\documentclass[copyright,creativecommons]{eptcs}
\providecommand{\event}{Types 2009} 

\usepackage{color}
\usepackage{graphicx}
\usepackage{amssymb}
\usepackage{fancyvrb}
\usepackage{ifpdf}
\usepackage{listings}
\usepackage{xypic}
\usepackage{bussproofs}

\hypersetup{plainpages=false, pdfborderstyle={/S/U/W 0}}

\renewcommand{\verb}{\lstinline}

\newtheorem{example}{Example}
\begin{document}

\title{Superposition as a logical glue}

\def\titlerunning{Superposition as a logical glue}
\def\authorrunning{A. Asperti and E. Tassi}

\author{Andrea Asperti
  \institute{Department of Computer Science, University of Bologna}
  \email{asperti@cs.unibo.it}
\and Enrico Tassi
 \institute{Microsoft Research-INRIA Joint Centre}
 \email{enrico.tassi@inria.fr}
}

\date{}
\maketitle

\begin{abstract}
The typical mathematical language systematically exploits 
notational and logical abuses whose resolution requires not just
the knowledge of domain specific notation and conventions, but
not trivial skills in the given mathematical discipline. 
A large part of this background knowledge is
expressed in form of equalities and isomorphisms, allowing mathematicians
to freely move between different incarnations of the same entity
without even mentioning the transformation. Providing ITP-systems with 
similar capabilities seems to be a major way to improve their intelligence,
and to ease the communication between the user and the machine. 
The present paper discusses our experience of integration of a superposition
calculus within the Matita interactive prover. Superposition provides the key
ingredient for a ``smart'' application tactic that allows the user to disregard
many tedious details otherwise needed to convince the system that his 
reasoning step is indeed correct.
We also show how this kind of automation, named \emph{small scale}, can serve
as the building block for the more general, \emph{large scale}, case, allowing
a smooth integration of equational reasoning with backward-based proof
searching procedures.
\end{abstract}

\section{Introduction}
One usually thinks of Mathematics as a precise discipline, 
often confusing mathematical rigor with logical formality.
In fact, most mathematics is simply too informal to be directly handled 
by the logical and algebraic means offered by interactive or automated 
theorem provers.
The typical mathematical discourse systematically exploits symbol 
overloading and notational abuses that can hardly be understood by 
automatic devices without substantial help from the user side,
that is one of the reasons why formal encoding is so expensive 
and frustrating.
The crucial point is that the intrinsic ambiguity of the mathematical 
vernacular can only be resolved by a sufficiently
contextual interpretation, requiring not just the knowledge of its 
specific notation and conventions, but nontrivial skills in the given 
mathematical discipline. A large part of this background knowledge is
expressed in form of equalities and isomorphisms, allowing a mathematician
to freely move between different incarnations (intensions) of the same entity
without even mentioning the transformation. Providing ITP-systems with 
similar capabilities seems to be a major way to improve their intelligence,
and to ease the communication between the user and the machine. In the
present paper, we discuss our experience of integration of a superposition 
calculus within the Matita interactive prover, providing in particular 
a very flexible, ``smart'' application tactic, and a simple, innovative 
approach to automation. 

The need for a stronger integration between fully automatic
(resolution) provers and interactive ones, and more generally for
a stronger automation support in proof assistants is a major 
challenge (see e.g. \cite{harrison-book})
and many efforts have been already done in this
direction: for instance, KIV has been integrated with the tableau 
prover $3T^AP$ \cite{AB98}; HOL has been integrated with various 
first order provers, such as 
Gandalf \cite{Hurd99} and Metis \cite{metis}; 
Coq has been integrated with Bliksem \cite{BHN02}; 
Isabelle was first
integrated with a purpose-built prover \cite{blast} and more recently with
Vampire \cite{MQP06}. 

We share most of the principles guiding these
efforts, and in particular the need to refer to a large library of 
known lemmas, and the goal to deliver a checkable proof, in conformity
with the trusted kernel philosophy (sometimes referred to as De Bruijn 
principle) inspiring most interactive provers.

However, there are two 
different uses of automation that have different requirements
and possibly deserve different approaches and solutions. 

The first one (small scale automation) 
is to relieve the user from the burden to fill in 
relatively trivial steps, by automatically completing the missing 
gaps. This kind of automation must be {\em fast}, {\em robust}
and sufficiently {\em predictable}, in the sense that the automation
procedure should not miss simple solutions when they exist. 
In general, in this case, the user is not interested to read
back the proof and there is no point in trying to transform it
in a human readable format. 

The second use of automation (large scale automation)
is to really help the user in the process
of {\em devising} the proof. In spite of all the progresses in this
field, a fully automatic approach still looks highly problematic: a 
more promising approach seems to be that of improving the 
cooperation between human and machine, and in particular of
making a better profit of the machine's  combinatorial capabilities.
At present, ``interaction'' in ITP systems is essentially restricted to
a master-slave command execution loop, that frustrates the computational
power of machines. A better repartition of the work could consist in
leaving to the user the most intelligent tasks, such as identifying 
the key lemmas, proof principles and intermediate results of interest
for the proof, assigning to the system the burden of composing them
into a coherent proof (eventually exploiting a huge library of known
results). The user has hence the responsibility to cut the search 
space, while the machine is supposed to automatically and
systematically explore it (this was also our guiding idea behind
the design of the automation driver of Matita version 0.5.x 
\cite{auto-driver}). Another recent system close to this conception is 
$\Omega$mega \cite{omega}: in this case the user drives the 
search by means of proof plans, invoking external reasoners 
to fullfill them. 

For large scale automation, we could bear
to run time consuming jobs, possibly working for hours in the background. 
The result is completely unpredictable, and probably unstable.
The system should produce a proof in a format as readable as possible, 
since the user is surely interested to inspect it himself, apart from
pasting it into the proof script (if proof searching is expensive
we probably wish to avoid running it over and over 
every time we recompile the script, or at least to have 
the possibility to choose between these two alternatives).
The heuristic, unstable nature of complex automation procedures, combined
with the verbose nature of fully formal proofs, naturally suggest 
the idea of producing ``proof traces'' as a compact, readable and
reproducible output of automation devices. 

This is the point where the two kinds of automation recombine together,
since the execution of proof traces precisely requires
small scale automation capabilities. Our point is that a good part
of these capabilities are fulfilled by reasoning up to equalities,
providing the {\em connective glue} that constitutes most
of that background knowledge tacitly, almost unconsciously used in
the typical mathematical reasoning. Superposition provides a
natural support for reasoning {\em modulo} a congruence on 
propositions, implementing ideas similar to \cite{modulo}, 
and providing a flexible and powerful tool for small scale
automation.

One of the components of the Matita interactive theorem prover 
is a state of the art, first order, untyped superposition algorithm, 
able to compete with the best tools currently 
available: in particular, our system scored in fourth position 
in the unit equality division at the 22nd CADE ATP System Competition, 
beating a glorious 
system such as Otter, and being awarded as the best 
new entrant tool \cite{Sutcliffe09}. 
Note in particular that 
Matita is entirely written in a functional language (OCaml), while most 
ATP system (with the relevant exception of Metis, that was however 
beaten by Matita) are written in imperative code.

In this paper, we shall provide a theoretical and
architectural description - as self contained as possible -
of this tool (Section \ref{sec:superposition}); 
hence we shall discuss its integration
with Matita (Section \ref{sec:integration}), and show some 
of its applications, mostly aimed to improve the intelligence 
of commands (smart application,
Section \ref{sec:applyS}) and the overall automation of the system 
(Section \ref{sec:auto}).

\section{Superposition}\label{sec:superposition}
Techniques for equational reasoning are a key component in many 
automated theorem provers and interactive proof and verification 
systems \cite{BG94,paramodulation,equality-handbook}. 
The main deductive mechanism is a {\em completion}
technique \cite{Knuth-Bendix} attempting to transform a given set of 
equations into a confluent rewriting system so that any two terms 
are equal if and only if they have identical normal forms. 
Not every equational theory can be presented as such a confluent
rewriting system, but you may progressively approximate it 
by means of a refutationally complete method called 
{\em ordered completion}.
The deductive inference rule used in completion procedures is 
{\em superposition} 
which consists of first unifying one side of one equation with a subterm 
of another, and hence rewriting it with the other side;
the selection of the two terms to be unified is guided by a given 
term ordering, which imposes certain restrictions on inferences, with the
major benefit to prune the search space. All results in this section
are known, and we only report them for the sake of completeness.

\subsection{Preliminaries}
Let $\mathcal{F}$ bet a countable alphabet of functional symbols, and 
$\mathcal{V}$ a countable alphabet of variables. 
We denote with $\mathcal{T}(\mathcal{F},\mathcal{V})$  the set of terms 
over $F$ with variables in $V$. A term 
$t\in \mathcal{T}(\mathcal{F},\mathcal{V})$ is either a 0-arity element of 
$\mathcal{F}$ (constant), an element of $\mathcal{V}$ (variable), 
or an expression of the form $f(t_1, \dots, t_n)$ where $f$ is a element of 
$\mathcal{F}$ of arity $n$ and $t_1, \dots, t_n$ are terms.

Let $s$ and $r$ be two terms. 
$s|_p$ denotes the subterm of $s$ at position $p$ and $s[r]_p$
denotes the term $s$ where the subterm at position $p$ has been 
replaced by $r$.

A substitution is a mapping from variables to terms. 
Two terms $s$ and $t$ are unifiable if there exists a substitution
$\sigma$ such that $s\sigma = t\sigma$. Moreover, in the previous case,
$\sigma$ is called a most general unifier (mgu) of $s$ and $t$ if 
for all substitution $\theta$ such that
$s\theta = t\theta$, there exists a substitution $\tau$ which satisfies
$\theta = \tau \circ \sigma$. 

A literal is either an abstract predicate (represented by a term), 
or an equality between two terms. A clause $\Gamma \vdash \Delta$ 
is a pair of multisets of literals: the negative literals $\Gamma$,
and the positive
ones $\Delta$. If $\Gamma = \emptyset$ (resp. $\Delta = \emptyset$), 
the clause is said to be positive (resp. negative).
A Horn clause is a clause with at most one positive literal. 
A unit clause is a clause
composed of a single literal. A unit equality is a unit clause 
where the literal is an equality. 

\subsection{Term orderings and Inference rules}
A strict ordering $\prec$ over $\mathcal{T}(\mathcal{F},\mathcal{V})$
is a transitive and irreflexive (possibly
partial) binary relation. An ordering is {\em stable} under substitution
if $s \prec t$ implies $s\sigma \prec t\sigma$ for all terms $t, s$ and
substitutions $\sigma$. A well founded monotonic ordering stable under
substitution is called {\em reduction ordering} (see  \cite{Dershowitz82}).
The intuition behind the use of reduction orderings for 
limiting the combinatorial explosion of the number of equations during
inference, is to only rewrite big terms to smaller ones.

\begin{description}
\item[superposition left] This rule defines backward reasoning steps. The
equational fact $l = r$ is combined with the goal $t_1 = t_2$ obtaining a
the goal $(t_1[r]_p = t_2)\sigma$.
  \begin{displaymath}
    \frac{
      \vdash l = r \quad\quad t_1 = t_2 \vdash
    }{
      (t_1[r]_p = t_2 \vdash)\sigma
    }
  \end{displaymath}
  if $\sigma = mgu(l, {t_1}|_p)$, $t_1|_p$ is not a variable, $l\sigma
  \not\preceq r\sigma$ and $t_1\sigma \not\preceq t_2\sigma$;\\

\item[superposition right] This rule defines forward reasoning steps. The two
equational facts $l = r$ and $t_1 = t_2$ obtaining a new fact
$(t_1[r]_p = t_2)\sigma$.
  \begin{displaymath}
    \frac{
      \vdash l = r \quad\quad \vdash t_1 = t_2
    }{
      (\vdash t_1[r]_p = t_2)\sigma
    }
  \end{displaymath}
  if $\sigma = mgu(l, {t_1}|_p)$, $t_1|_p$ is not a variable, $l\sigma
  \not\preceq r\sigma$ and $t_1\sigma \not\preceq t_2\sigma$;\\

\item[equality resolution] This is the rule that ends the proof search.
  \begin{displaymath}
    \frac{
      t_1 = t_2 \vdash
    }{
      \Box
    }
  \end{displaymath}
  if there exists $\sigma = mgu(t_1, t_2)$.
\end{description}

\subsection{Simplification rules}
For efficiency reasons, the calculus must be integrated with a few
additional optimization rules, the most important one being
demodulation (\cite{demodulation}):

\begin{description}
\item[subsumption] This rule allows to identify and drop clauses that happen to
be the instance of more general ones, and are thus superflous.
  \begin{displaymath}
    S \cup \{C, D\} \Rightarrow S \cup \{C\}
  \end{displaymath}
  if $C$ \emph{subsumes} $D$, i.e. if there exists a substitution $\sigma$
  such that $C\sigma \equiv D$.\smallskip

\item[tautology elimination] This rule eliminates equational facts that
are provable with the \textbf{equality resolution} rule and are thus superflous.
  \begin{displaymath}
    S \cup \{\vdash t = t\} \Rightarrow S.
  \end{displaymath}

\item[demodulation] This rule aims at reducing the size of the clauses involved
in the proof search, speeding up all operations whose complexity is determined
by the size of the terms involved. The intuitive idea is to consider clauses
modulo know equational facts and record only their smaller representative.
  \vspace*{-.1cm}
  \begin{displaymath}
    S \cup \{\vdash l = r, C\} \Rightarrow S \cup
    \{\vdash l = r, C[r\sigma]_p\},
  \end{displaymath}
  if $l\sigma \equiv C|_p$ and $l\sigma \succ r\sigma$.
\end{description}

\subsection{The main algorithm}
\label{sec:main_algorithm}
\begin{figure}[htp]
\begin{center}
\includegraphics[width=0.50\textwidth]{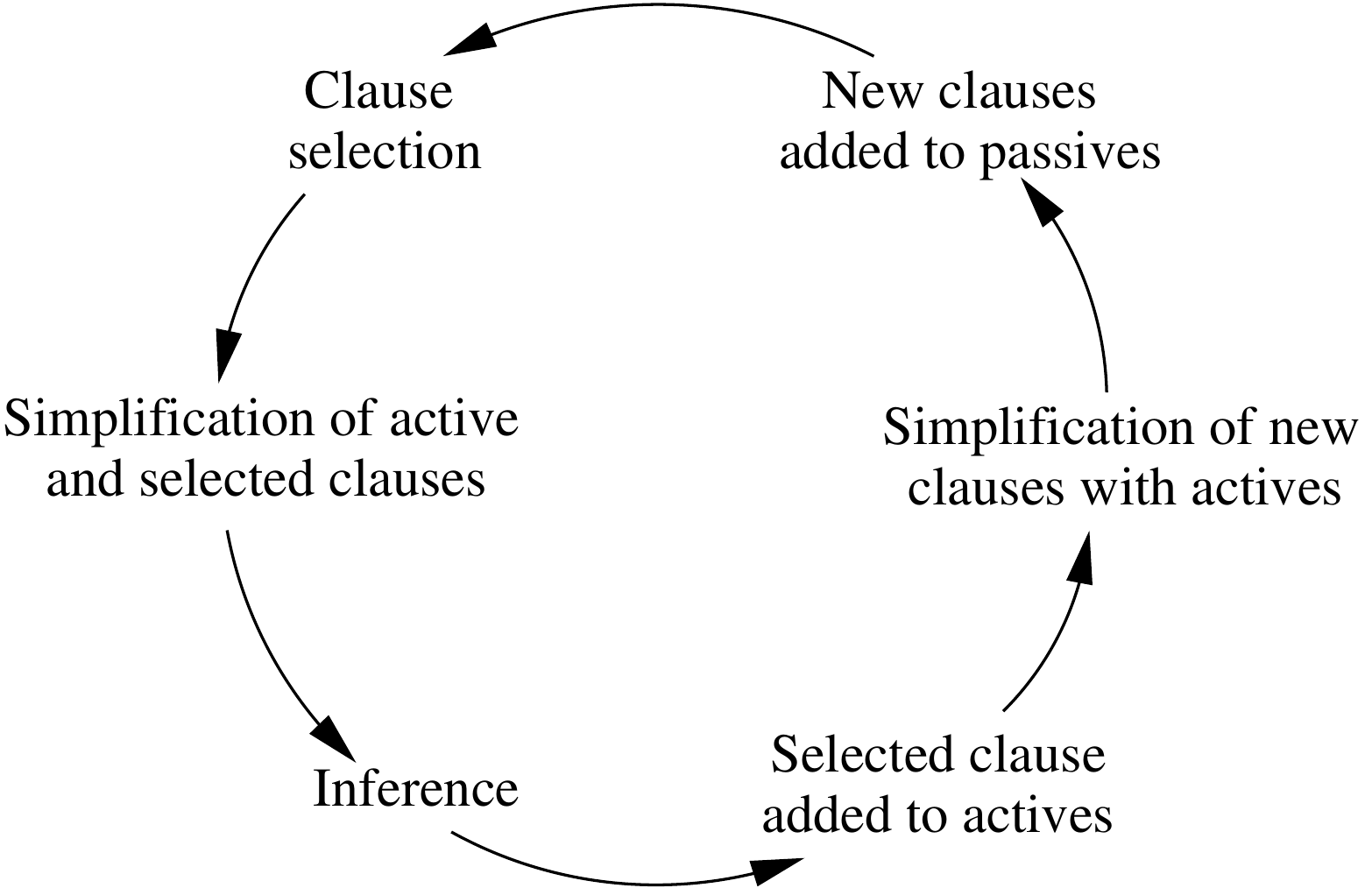}
\caption{given-clause algorithm\label{mainloop}}
\end{center}
\end{figure}

To avoid combining the same clauses twice, it is convenient to keep
them in two sets, that are traditionally called {\em active} and
{\em passive}. The general invariant is that clauses in the active 
sets have been already composed together in all possible ways. 
A step consists in selecting some clauses from the passive set, 
add them to the active set, compose them with the current active set 
- and thus with themselves - (inference), 
and finally add the newly generated clauses to the passive set 
(possibly after a simplification). 
A natural strategy would consist in selecting the whole passive set
at each iteration, realizing a sort of breadthfirst strategy. 
The advantage of this strategy is that it is very predictable, and 
hence particularly easy to debug. Unfortunately the number of new
equations generated at each step grows extremely fast, practically
preventing to iterate the main loop more than a few steps.  
To avoid this problem, the opposite solution is usually adopted, 
consisting in selecting just {\em one} passive equation at each step.
The equation is selected according to suitable heuristics (size, 
goal similarity, and so on), usually comprising some fairness 
criterion to ensure completeness (we must ensure that any passive
equation will be selected, soon or later). 
This approach 
is called the \emph{given-clause algorithm} (figure \ref{mainloop}), 
and it is the procedure used (with some variations)
by all modern theorem provers (see e.g. \cite{vampire_annals}).
The advantage of this method
is that the passive set grows much slower, allowing a more focused
and deeper inspection of the search space. The drawback is that 
the algorithm becomes extremely sensitive to the selection heuristic,
leading to more unpredictable behaviour. 

\smallskip
In order to get a high performance tool, the given clause algorithm has 
to be tuned and optimized in several ways. The critical areas are:
\begin{itemize}
\item Data structures and code optimization
\item Orderings used to orientate rewriting rules
\item Selection strategy
\item Demodulation
\end{itemize}
We are currently using relatively 
simple data structures (discrimination 
trees \cite{McCune}) for term indexing, 
but we plan to exploit in the near future more specific data structures 
(such as substitution \cite{Graf95} or context trees \cite{codedcontexttrees}).

On complex problems (e.g. problems in the TPTP library with rating
greater then $0.30$) the choice of a good ordering for inference rules 
is of critical importance. 
As we already mentioned, we have implemented several orderings, 
comprising standard Knuth-Bendix (KBO), non recursive Knuth-Bendix
(NRKBO), lexicographic path ordering (LPO) and recursive path ordering (RPO).
The best suited ordering heavily depends on the kind of problem, 
and is hard to predict\footnote{Our approach to the CADE ATP System Competition
was to run in parallel different processes with different orderings.}.


Luckily, on simpler problems (of the kind required for small
scale automation) the given-clause algorithm is less sensitive
to the term-ordering, and any of them usually produce a solution in a 
reasonable amount of time. 

The selection strategy currently implemented by Matita is a based on
combination of age and weight. The weight is a positive integer that 
provides an estimation of the ``complexity'' of the
clause, and is tightly related to the number of occurrences of symbols 
in it.

Another important issue for performance is demodulation: the given clause
algorithm spends most of its time (up to 80\%) in simplification, hence
any improvement in this part of the code has a deep impact on performance.
However, while reductions strategies, sharing issues and abstract machines
have been extensively investigated for lambda calculus (and in general
for left linear systems) less is known for general first 
order rewriting systems. 
In particular, while an innermost (eager) 
reduction strategy seem to work generally better than an outermost 
one (especially when combined with lexicographic path ordering), one 
could easily create examples showing an opposite behaviour (even 
supposing to always reduce needed redexes). 


Although we did not want to focus too much on developing specific 
heuristics, two widespread techniques, not yet implemented, would 
still be of great interest. The
first one is Limited Resource Strategy \cite{LRS}, which basically allows 
the procedure to skip some inference steps if the resulting clauses 
are unlikely to be processed,  because of a lack of time or memory. 
The other promising technique is indexing
modulo associativity and commutativity \cite{assoccommut}, which is often 
heavily used when working on algebraic structures.

\section{Integrating superposition with Matita}
\label{sec:integration}

\subsection{Library management}
A simple possibility for integrating superposition with Matita
is simply to solve goals assuming as initial passive set all
equational facts in the library (plus the equations in the
local context). The main drawback of this approach is that
passive equations would be selected slowly, and in a quite
{\em repetitive way} every time a new problem is met. In fact, superposition
right, as any forward operation, only concern {\em facts}, and 
apart from the local hypothesis, most of these facts are known
in advance. This suggest that, in ITP systems, forward operations
should be processed, as much as possible, off line; 
but then we have to face the
dual problem, namely to avoid an unnecessary proliferation of 
results, polluting the library (and the memory) with trivialities. 
The compromise adopted in Matita was suggested
by the observation that, in a given-clause algorithm, 
{\em selection} is a conspicuous operation
requiring an intelligent choice; but all theorems in the library
are indeed already a ``selected'' subset (otherwise, there would be
no point to record them). In other words, the idea is 
to use the unit equalities in the library not as initial passive set, 
but as the {\em active} one. This means that every time a new 
equality is added to the library it also goes through
one cycle of the given-clause algorithm as if it was the newly
selected passive equation: it is composed (after simplification)
with all existing active equations (that is, up to simplifications,
all previously proved equalities), and the newly created equations
are added to the passive list\footnote{This approach is particularly
important in view of the fact that, typically, the passive set is not
even used for demodulation.} This way, we have a natural, 
simple but traceable syntax to drive the saturation process: it is 
enough to explicitly list in the library the selected equation.
At the same time, this approach reduces the verbosity of the library, since
trivial results generated by superposition in the passive list
may be used without the need to declare (and name) them explicitly.

\subsection{Input/Output}

The communication between Matita and the superposition tool is
not precise. As we already said, our superposition algorithm is
first order and untyped; instead of attempting a complex encoding
of the Calculus of Inductive Constructions (CIC) in first order logic 
(that is the approach adopted e.g. in \cite{MP08}), we prefer to
use a naive, but efficient translation, possibly losing information.
We shall then try to automatically reconstruct the missing information
during proof reconstruction, exploiting the sophisticated inference
capability of the Matita {\em refiner} \cite{hints}. 
As a consequence, automation is a best effort service: 
not only it may obviously fail to produce a proof, but sometimes
it could produce an argument that the system is not able to refine correctly
(independently from the fact if the delivered proof was ``correct''
or less).

Although there is no particular problem to implement a typed 
superposition algorithm, or even embedding types as first order
terms (in more or less naive ways, according to the way 
we wish to take convertibility into account), for performance 
reasons we decided to work with completely untyped terms.
In particular, equations $r =_T s$ of the calculus of constructions
are translated to first order equations by merely following the
applicative structure of $r$ and $s$, and translating 
any other subterm into an opaque constant. The type $T$ of the equation
is recorded, but we are not supposed to be able to compute types
for subterms.

Since all equations are combined together via superposition rules, there is
a (modest) risk of producing ``ill-typed'' terms.
Consider for instance the superposition left rule (the reasoning is similar
for the other rules)
  \begin{displaymath}
    \frac{
      \vdash l = r \quad\quad t_1 = t_2 \vdash
    }{
      (t_1[r]_p = t_2 \vdash)\sigma
    }
  \end{displaymath}
where $\sigma = mgu(l, {t_1}|_p)$ and $l\sigma \not\preceq r\sigma$.
The risk is that $t_1|_p$ has 
a different type from $l$, hence resulting into an illegal rewriting
step. Note however that $l$ and $r$ are usually rigid terms, whose
type is uniquely determined by the outermost symbol. Moreover, 
$t_1|_p$ cannot be a variable, hence they must share this outermost
symbol. If $l$ is not rigid, it is usually a variable $x$ and if 
$x \in r$ (like e.g. in $x=x+0$) we have (in most orderings)
$l \preceq r$ that again rules out rewriting in the wrong direction.

This leads us to the following notion of {\em admissibility}. 
We say that an applicative term $f(x_1,\dots,x_n)$ is {\em implicitly
typed} if its type is uniquely determined by the type of $f$. 
We say that an equation $l = r$ is admissible if both $l$ and $r$
are implicitly typed, or $l \preceq r$ and $r$ is implicitly typed.
If an equation is not admissible, we forbid to take it into account
for superposition. 


\subsection{(Re)construction of the proof term}\label{sec:proofs}

Reading back a superposition proof inside any interactive prover
is a relatively simple operation (just requiring rewriting), 
and one of the reasons for sticking to this fragment. 
In the superposition module, each proof step in encoded as a 
tuple
\begin{verbatim}
  Step of rule * int * int * direction * position * substitution
\end{verbatim}
where \verb+rule+ is the kind of rule which has been applied,
the two integers are the two $id's$ of the composing equations
(referring to a ``bag'' of unit clauses),
\verb+direction+ is the direction the second equation is applied
to the first one, \verb+position+ is a path inside the rewritten term and
finally \verb+substitution+ is the mgu required for the rewriting step.
\begin{figure}[htp]
\begin{center}
\includegraphics[width=0.5\textwidth]{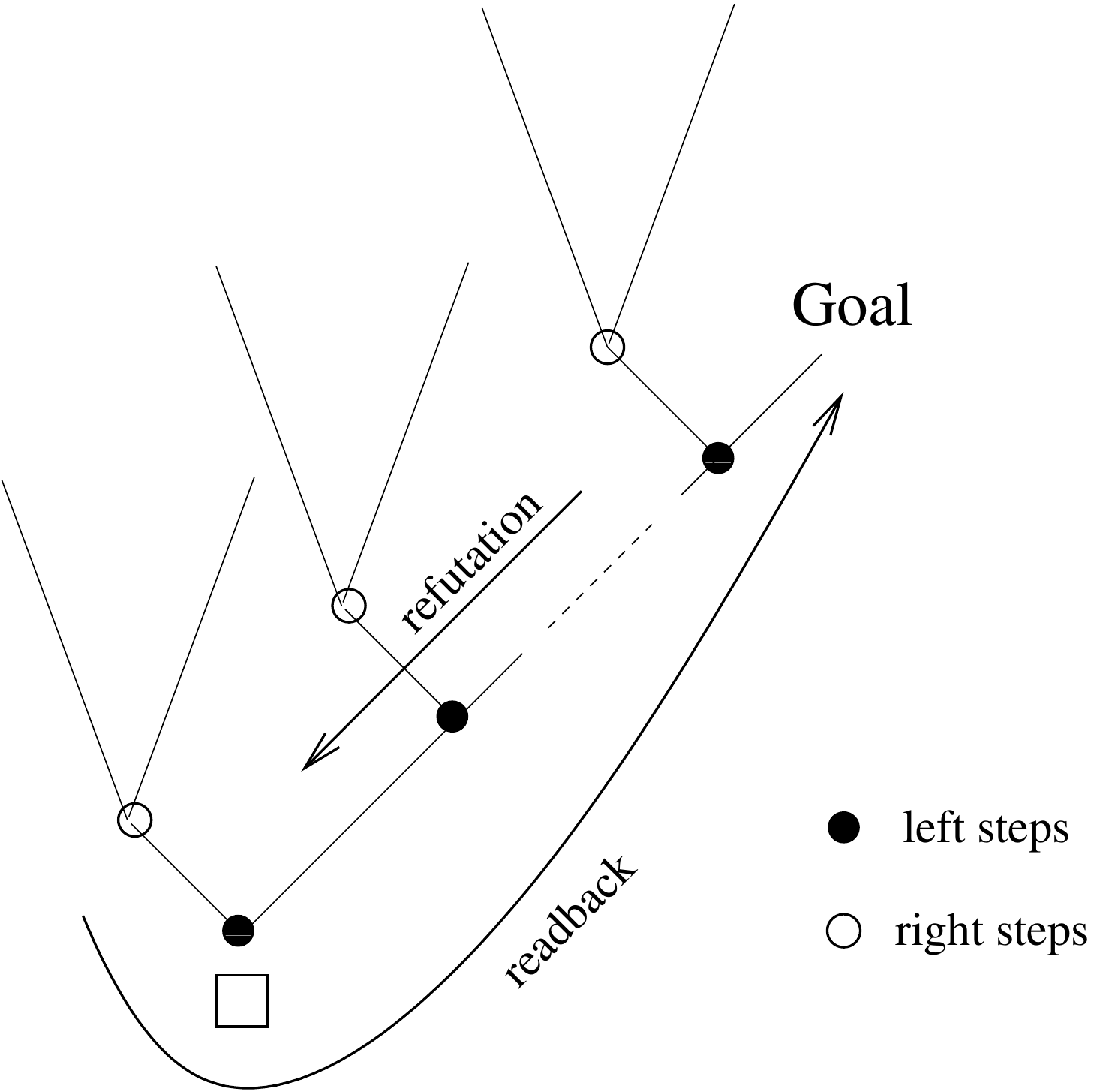}
\caption{given-clause algorithm\label{readback}}
\end{center}
\end{figure}
The proof has the shape depicted in Figure \ref{readback}, where all 
superposition left steps are on the rightmost spine leading from the
Goal to the empty clause. Superpostion right steps are
forward rewriting operations and their translation is straightforward;
on the other side, superposition left steps must be reverted in order to build
a direct proof of the (suitably instantiated) goal from its refutation. 

\noindent
Formally, let's call \verb+eq_ind+ the higher order
rewriting step
\[
eq\_ind: \forall A: \textsc{Type}. \forall x:A. \forall P: A \to
  \textsc{Prop}.P~x \to \forall y: A. x = y \to P~y
\]
Let us consider a superposition right step 
\[
  \frac{
      \vdash l =_A r \quad\quad \vdash t =_B s
    }{
      \vdash t[r]_p\sigma =_B s\sigma
    }
\]
If $h:l =_A r$ and $k:t =_B s$ then 
  \[eq\_ind ~A~l\sigma~ (\lambda x:A.t[x]_p =_B s)\sigma~k\sigma~
       r\sigma~h\sigma : t[r]_p\sigma =_B s\sigma \]
Conversely, given a superposition left step 
\[
    \frac{
      \vdash l =_A r \quad\quad \alpha:t =_B s\vdash  
    }{
       t[r]_p\sigma =_B s\sigma \vdash  
    }
\]
if $h:r =_A l$ and $k: t[r]_p\sigma =_B s\sigma$ then
\[eq\_ind ~A~r\sigma~ (\lambda x:A.t[x]_p =_B s)\sigma~
      k~l\sigma~h\sigma : t\sigma =_B s\sigma
\]


To generate a CIC proof term, clauses are topologically sorted
w.r.t. their dependendices (to respect scoping), 
their free variables are explicitly quantified, and nested let-in 
patterns are used to build the proof.
A delicate point of the translation is closing each clause
w.r.t. its free variables, since we should infer a type for them.
The simplest solution is to generate so called ``implicit'' arguments
leaving to the Matita {\em refiner}~\cite{hints}, the burden of guessing them.
For instance, superposing $plusC: x + y = y \underline{+} x$ with 
$plusA: x + (y + z) \stackrel{\leftarrow}{=} (x + y) + z$ at the underlined 
position and in the given direction
gives rise to the following piece of code, where question marks stands
for implicit arguments:
 
\begin{lstlisting}
let clause_59:
   $\forall x183: ?.$
    $\forall x184: ?.$
     $\forall x185: ?.$
       $x183 + (x184 + x185)) = x184 + (x185 + x183)$
   $:= \lambda x183:?.$ 
       $\lambda x184:?.$
        $\lambda x185:?.$
          eq_ind$nat\; ((x184 + x185) + x183)$
            $(\lambda x:nat. x183 + (x184 + x185) = x)$
            (plusC $x183 (x184 + x185)) (x184 + (x185 + x183)$)
            (plusA $x184 x185 x183$) in
...
\end{lstlisting}
%

\section{Smart applications}
\label{sec:applyS}
The first interesting application of superposition (apart its
use for solving equational goals), is the implementation of a more
flexible application tactic. As a matter of fact, one of the most 
annoying aspects of formal development is the need of transforming 
notions to match 
a given, existing result. As we already said, most of these 
transformations are completely transparent to the typical mathematical
discourse, and we would like to obtain a similar behaviour in interactive
provers.

Given a goal $G$ and a theorem t: $\Gamma \to A$, the goal is to try
to match $A$ with $G$ up to the available equational knowledge base, in
order to apply $t$. We call it, the {\em smart application} of $t$ to 
$G$. 

We use superposition in the most direct way, exploiting on
one side the higher-order features of CIC, and on the other
the fact that the translation to first order terms does
not make any difference between predicates and functions:
we simply generate a
goal $A = G$ and pass it to the superposition tool (actually, 
it was precisely this kind of operation that motivated our original
interest in superposition). If a proof is found, $G$ is transformed
into $A$ by rewriting and $t$ is then 
normally applied.

Superposition, addressing a typically undecidable problem, 
can easily diverge, while we would like to have a reasonably 
fast answer to the smart application invocation, as for any other 
tactic of the system. We could simply
add a timeout, but we prefer to take a different, more predictable 
approach. As we already said, the overall idea is that superposition
right steps - realising the {\em saturation} of the equational
theory - should be thought of as off line operations. Hence, at run
time, we should conceptually work as if we had a {\em 
confluent} rewriting system, and the only operation worth to do
is {\em narrowing} (that is, left superposition steps). Narrowing
too can be undecidable, hence we fix a given number of narrowing 
operations to apply to each goal (where the new goal instances generated at
each step are treated in parallel). The number of narrowing steps 
can be fixed by the user, but a really small number is usually
enough to solve the problem if a solution exists. 

\begin{example}
\label{example:smart}
Suppose we wish to prove that the successor function
is $\le$-reflecting,
namely
\[(*)\hspace{.5cm}\forall n,m. S n \le S m \to n \le m\]
Suppose we already proved that the predecessor function is monotonic:
\[monotonic\_pred: \forall n,m. n \le m \to pred\; n \le pred\; m\]
We would like to merely ``apply'' the latter to prove the former.
Unfortunately, this would not work, since there is no way to match
$pred\; X \le pred\; Y$ versus $n \le m$, unless {\em narrowing} the 
former. By superposing twice with the equation 
$\forall n. pred (S n) = n$ we 
immediately solve our matching problem via the substitution
$\{X := S n, Y := S m\}$.  Hence, the smart application of
$monotonic\_pred$ to the goal $n \le m$ succeeds, opening the new
goal $S n \le S m$ that is the assumption in $(*)$.
\end{example}

\begin{example}
\label{example:smart8}
Let us use the notation $A[B/i]$ to express the substitution of
$B$ for the $i-th$ free variable in $A$. The substitution lemma
says that for all $k,i$
\[A[B/i][C/i+k] = A[C/S(k+i)][B[C/k]/i]\]
(where $S$ is the successor function).
The idea is to prove the substitution lemma by structural induction
over $A$. Suppose now $A$ is a binder, e.g. a lambda term $FUN(M)$ 
where $M$ is the body of the function. 
The definition of substitution tells us that
\[FUN(M)[B/i] = FUN(M[B/i+1])\]
Hence, after normalization and elimination of congruent terms,
we are left to prove\footnote{We added some artificial 
parenthesis to the terms to emphasize the left associativity of
plus.}
\[(M[B/i+1][C/(k+i)+1] = Fun(M[C/S((k+i)+1)][B[C/k]/i+1]\]
under the inductive hypothesis
\[Hind: \forall j.M[B/i][C/k+j] = M[C/S(k+j)][B[C/k]/j]\]
It is evident that it is enough to instantiate $j$ with $i+1$
but in order to unify $(k+i)+1$ with $k+?_j$ we have to use the
associativity law for the sum! Hence the smart application of
$Hind$ succeeds where the normal application would fail.
\end{example}

\section{The auto tactic}
\label{sec:auto}
By itself, smart application is less interesting than expected. 
The point is that, compared to the effort of {\em finding} the 
``right'' theorem $t$ in the library, the work of transforming the goal 
to match the conclusion is a boring, but minor task. 

What is really interesting, instead, is the possibility to combine
smart application with a goal-oriented proof searching
technique, to achieve a cheap, simple but surprisingly effective 
management of equality.  

According to our philosophy, forward operations in ITP systems 
should be performed off line, and explicitly or implicitly
recorded in the library (if a forward step is {\em really} useful
in some context, it is likely to be useful in other, similar 
contexts as well, hence it is a very good candidate to explicitly 
appear in the library). For this reason, the Matita automation 
tool is backward-based (backward operations act on the goal, that
is only known at run time), essentially trying to build a proof
by a repetitive application of tactics. The proof we are looking for is not
in normal form: in fact, the most relevant tactic is application, 
and the automation tool is supposed to explore the library for 
all known results matching the current goal. In this respect, 
automation resembles a prolog-like program, and we use a traditional
depth-first strategy (with bounded, user configurable depth) 
to explore the proof space. 
The main optimizations\footnote{All these optimizations destroy the so called
procedural interpretation of logic programs, and received 
very little attention by the logic programming community.}
 implemented are the following:
\begin{description}
\item[goal clustering] a {\em cluster} in a set of (conjunctive)
goals $\Delta = g_1,\dots,g_n$ is a minimal subset closed w.r.t. its
free variables: any variable appearing in a goal of the cluster
can only appear in other goals of the same cluster. Clusters 
obviously form a partition of the original set; 
their interest is that the processing of different clusters 
can be separated by {\em green cuts};
\item[loop detection] if a goal $\Delta$ generates 
another goal $\Delta'$ subsumed by the former, the proof branch can
be pruned\footnote{At present this is only implemented in case $\Delta'$
is a single literal.}(if we find a proof for $\Delta'$ it works 
for $\Delta$ as well - recall that variables in goals are 
existentially quantified). 
\end{description}
We also plan to implement a failure cache (indexed by the failure
depth); instead, the advantage of caching successes
looks much more questionable (either we pre-compute the
whole success set, requiring a different proof searching
strategy, or we easily end-up duplicating solutions).

Smart application can be easily integrated in our automated
proof searching tactic. Per se, due to the severe constraints 
imposed on superposition, smart application is not much slower than
normal application. The real problem is the brutal explosion in the 
number of candidates. With normal application, using good data
structures for indexing the universe of known results (we use
discrimination trees \cite{McCune}), we are able to retrieve, for each goal, 
a relatively small number of candidates. In the case of smart
application, any theorem predicating something ``similar'' to the goal
is a potential candidate. Our notion of similarity is particularly
weak: we look for any theorem whose conclusion shares with the goal 
(possibly up to reduction) the same top predicate. 

Note however that what really
matters from the complexity point of view is not the number
of candidates which are tried, but the number of them whose
application {\em succeeds}, giving rise to new branches in the
the search tree. Luckily, in general, smart application does not 
sensibly enlarge the number of applicable theorems, and the overall
complexity remains feasible, especially for small depths (3/4). 

\subsection{Proof traces}
Since most of the time is spent in searching the right theorems
composing the proof, a natural idea is to let the automation
tactic return a trace of the proof consisting of all
library results used to build the proof. 
We omit the local assumptions, all equations used by superposition, 
and to further reduce the verbosity of the trace,
we also omit all library {\em facts} (i.e. all results with no
hypothesis, hence appearing in leaf position inside the proof). 
The resulting set is passed as an optional ``by'' argument to
the auto tactic.

If the argument is present, the automation tactic would use
the set passed as an argument as candidates for smart applications, 
apart from at depth $0$, where facts in the whole library would be
taken into account. Local assumptions are always tried, too.

Using these simple proof traces automation becomes extremely fast,
and almost comparable to a fully expanded proof script. 


\subsubsection{Example}
This is a relatively complex example borrowed from the Matita standard
library (in particular, in a contribution regarding lifting
and substitution in DeBrujin notation). The goal to prove is
$k \le n-1$ under the assumption $H: j + k < n$, where $j,k$ and $n$ 
are natural numbers. The relation $n < m$ is definitionally equivalent
to $S n \le m$ where $S$ is the successor function. Note that
the successor function is extensionally equal to (but does not 
coincide with) the operation of adding 1, in the same way as the 
predecessor function is extensionally equal to (but does not coincide with) 
the operation of subtracting 1. Another delicate point is that 
the minus operation $x - y$ returns $0$ when $y > x$, so
$S(x - 1) = x$ only if $x > 0$. \\
The solution automatically found
by Matita is depicted in Figure \ref{example_glue} (the picture 
is better understood reading it from the bottom to the top): 
\begin{figure}[htp]
\begin{center}
\includegraphics[width=0.5\textwidth]{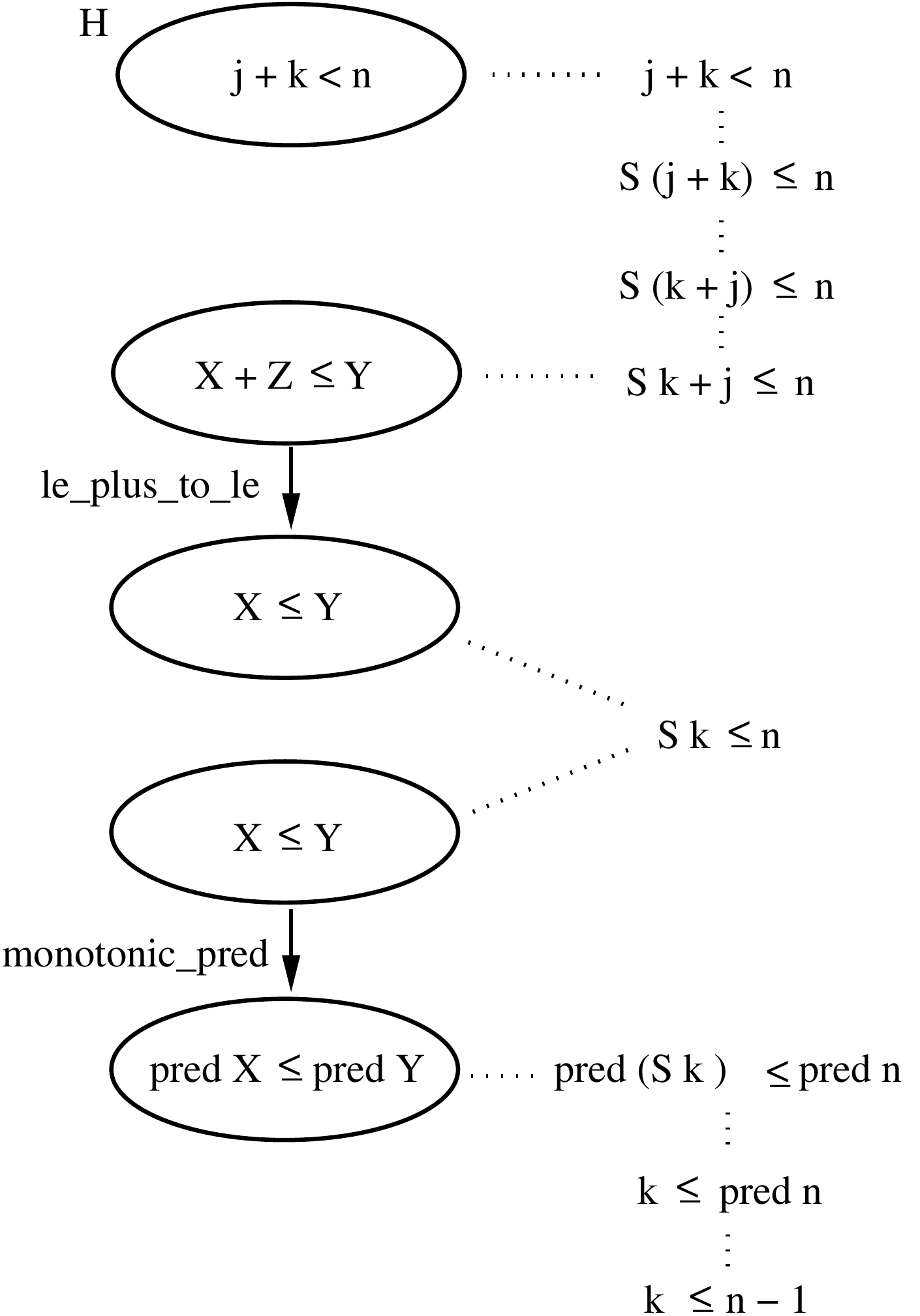}
\caption{Glueing together lemmas via rewriting\label{example_glue}}
\end{center}
\end{figure}
it first applies the monotonicity of the
predecessor function, passing from $k \le n-1$ to $S k \le n$; then
it applies the lemma
\[le\_plus\_to\_le: \forall n,m,a: n + a \le m \to n \le m\] 
obtaining the goal $k + Z \le n$, and finally applies the 
hypothesis $H$, instantiating $Z$ with $S\;j$. In order to do these
passages, the system exploits the following equivalences: $k = pred (S\;k)$, $n - 1 = pred\;n$, $S k \le n = k < n$,
$S (j + k) = (S\;j) + k = k + (S \;j)$; they are
the {\em logical glue} permitting to compose together the relevant
applicative steps (\verb+le_plus_to_le+ and \verb+monotonic_pred+, 
forming the {\em trace} of the proof). 

\subsubsection{Some statistics}
We are currently porting the old matita library (containing almost 
no automation) to the new Matita system. 
The following table compares the two
libraries on a fragment of about one hundred elementary arithmetical theorems.
The ``size'' is the dimension of the gzipped file in bytes.

\[
\begin{array}{c|c|c|c|c|}
                   & \mbox{lines}  & \mbox{size}   &  \mbox{size} & \mbox{compilation} \\
                   &               & \mbox{(whole)}&  \mbox{(proofs)} & \mbox{time}\\\hline
\mbox{no auto}     &    1139       &    5753     &    3433 (60\%) &     4.6s   \\\hline
\mbox{with auto}   &     627       &    3788     &    2027 (53\%) &    50.4s   \\\hline
\mbox{traces}      &     627       &    3982     &    2163 (54\%) &     5.3s   \\\hline
\end{array}
\]
The sensible increase of compilation time in presence of automation
was somehow expected: the leading idea of the paper is that we are 
ready to pay some extra execution time if this can reduce the 
encoding effort on the user side (provided it does not sensibly 
slow-down the system reactiveness to user commands in interactive 
sessions: note that the average execution time per theorem is about
0.35 seconds). Traces seem to provide a natural balance between performance
and verbosity.

\section{Conclusions}
\label{sec:conclusions}
In this paper we introduced a general methodology to address
the complex problem of automation in interactive provers. The main 
principles underlying our approach are the following:
\begin{enumerate}
\item there is an important distinction to be made between {\em small scale 
automation}, mostly meant to reduce the verbosity of the proof script
(resolution of trivial steps, verification of side conditions, 
automatic inference of missing information, etc.),
and {\em large scale automation} (problem solving): the problems and
requirements in the two cases are different, eventually
deserving different approaches and solutions;
\item a major component of small scale automation is the capability
to ``reason'' (apply logical rules and theorems) up to equalities,
covering most part of the background knowledge tacitly used in
the typical mathematical reasoning as an underlying connective
glue between logical steps (see Figure\ref{example_glue}); 
\item large scale automation must return a human readable and 
system executable proof trace; the trace must be simple, hence
its execution will eventually require small scale automation 
capabilites (independently of the choice of implementing or not 
large scale automation {\em on top} of small scale automation).
\end{enumerate}
The paper also describes the current state of the implementation of
this program inside the Matita interactive theorem prover.
In particular we presented the architectural design of the
superposition tool supporting equational reasoning, and its
{\em not so trivial} integration inside the Matita Interactive Theorem
Prover.

The tool is already highly performant (we scored in fourth position 
in the unit equality division at the 22nd CADE ATP System Competition), 
but many improvements can still be done for efficiency. 
In particular, more specialised data structures for indexes would 
hopefully give us a chance to scale up with the current best ATPs.

Another interesting research direction is to extend the management
of equality to setoid rewriting \cite{setoidrewriting}. 
Indeed, the current version of the superposition tool merely works
with an intensional equality, and it would be
interesting to try to figure out how to handle more general 
binary relations. The main problem is proof reconstruction, but
again it looks possible to exploit the sophisticated capabilities
of the Matita refiner \cite{hints} to automatically check the legality of
the rewriting operation (i.e. the monotonicity of the context
inside which rewriting has to be performed). 


While we are, at present, reasonably happy of the small scale 
automation capabilities of Matita, much work is left about
large scale automation. Our current approach tries to build
large scale automation {\em on top} of small scale automation
(e.g. substituting application by its smart version); this
approach is natural, especially in view of the generation of 
proof traces but, as we already observed, it is not the only 
possibility compatible with our methodology and alternative
solutions (requiring a tighter integration of small scale
automation {\em techniques} inside large scale functionalities)
are worth to be explored.

Finally, let us remark that proof traces are per se an interesting 
object worthty of further investigation (and, possibly, standardization),
in order to optimize the trade-off between efficiency and verbosity,
or to improve interoperability bewteen different systems.


\bibliographystyle{plain}
\bibliography{../BIBTEX/helm}

\begin{thebibliography}{10}
\providecommand{\bibitemdeclare}[2]{}
\providecommand{\urlprefix}{Available at }
\providecommand{\url}[1]{\texttt{#1}}
\providecommand{\href}[2]{\texttt{#2}}
\providecommand{\urlalt}[2]{\href{#1}{#2}}
\providecommand{\doi}[1]{doi:\urlalt{http://dx.doi.org/#1}{#1}}
\providecommand{\bibinfo}[2]{#2}

\bibitemdeclare{incollection}{AB98}
\bibitem{AB98}
\bibinfo{author}{Wolfgang Ahrendt}, \bibinfo{author}{Bernhard Beckert},
  \bibinfo{author}{Reiner H\"ahnle}, \bibinfo{author}{Wolfram Menzel},
  \bibinfo{author}{Wolfgang Reif}, \bibinfo{author}{Gerhard Schellhorn} \&
  \bibinfo{author}{Peter~H. Schmitt} (\bibinfo{year}{1998}):
  \emph{\bibinfo{title}{Integrating Automated and Interactive Theorem
  Proving}}.
\newblock In \bibinfo{editor}{Wolfgang Bibel} \& \bibinfo{editor}{Peter~H.
  Schmitt}, editors: {\sl \bibinfo{booktitle}{Automated Deduction --- A Basis
  for Applications}}. {\sl \bibinfo{series}{Applied Logic Series, No. 9}}
  \bibinfo{volume}{{II}: Systems and Implementation Techniques},
  \bibinfo{publisher}{Kluwer, Dordrecht}, pp. \bibinfo{pages}{97--116}.

\bibitemdeclare{inproceedings}{hints}
\bibitem{hints}
\bibinfo{author}{Andrea Asperti}, \bibinfo{author}{Wilmer Ricciotti},
  \bibinfo{author}{Claudio~Sacerdoti Coen} \& \bibinfo{author}{Enrico Tassi}
  (\bibinfo{year}{2009}): \emph{\bibinfo{title}{Hints in Unification}}.
\newblock In \bibinfo{editor}{Stefan Berghofer}, \bibinfo{editor}{Tobias
  Nipkow}, \bibinfo{editor}{Christian Urban} \& \bibinfo{editor}{Makarius
  Wenzel}, editors: {\sl \bibinfo{booktitle}{TPHOLs}}. {\sl
  \bibinfo{series}{Lecture Notes in Computer Science}} \bibinfo{volume}{5674},
  \bibinfo{publisher}{Springer}, pp. \bibinfo{pages}{84--98},
  \doi{10.1007/978-3-642-03359-9\_8}.

\bibitemdeclare{inproceedings}{auto-driver}
\bibitem{auto-driver}
\bibinfo{author}{Andrea Asperti} \& \bibinfo{author}{Enrico Tassi}
  (\bibinfo{year}{2009}): \emph{\bibinfo{title}{An interactive driver for goal
  directed proof strategies}}.
\newblock In: {\sl \bibinfo{booktitle}{Proc. of User Interfaces for Theorem
  Provers 2008. Montreal, CA, August 2008}}. {\sl \bibinfo{series}{ENTCS}}
  \bibinfo{volume}{226}, pp. \bibinfo{pages}{89--105},
  \doi{10.1016/j.entcs.2008.12.099}.

\bibitemdeclare{article}{BG94}
\bibitem{BG94}
\bibinfo{author}{Leo Bachmair} \& \bibinfo{author}{Harald Ganzinger}
  (\bibinfo{year}{1994}): \emph{\bibinfo{title}{Rewrite-Based Equational
  Theorem Proving with Selection and Simplification}}.
\newblock {\sl \bibinfo{journal}{J. Log. Comput.}}
  \bibinfo{volume}{4}(\bibinfo{number}{3}), pp. \bibinfo{pages}{217--247},
  \doi{10.1093/logcom/4.3.217}.

\bibitemdeclare{inproceedings}{omega}
\bibitem{omega}
\bibinfo{author}{Christoph Benzm{\"u}ller}, \bibinfo{author}{Armin Fiedler},
  \bibinfo{author}{Andreas Meier}, \bibinfo{author}{Martin Pollet} \&
  \bibinfo{author}{J{\"o}rg~H. Siekmann} (\bibinfo{year}{2006}):
  \emph{\bibinfo{title}{Omega}}.
\newblock In: {\sl \bibinfo{booktitle}{The Seventeen Provers of the World}}.
  {\sl \bibinfo{series}{Lecture Notes in Computer Science}}
  \bibinfo{volume}{3600}, \bibinfo{publisher}{Springer}, pp.
  \bibinfo{pages}{127--141}, \doi{10.1007/11542384\_17}.

\bibitemdeclare{article}{BHN02}
\bibitem{BHN02}
\bibinfo{author}{Marc Bezem}, \bibinfo{author}{Dimitri Hendriks} \&
  \bibinfo{author}{Hans de~Nivelle} (\bibinfo{year}{2002}):
  \emph{\bibinfo{title}{Automated Proof Construction in Type Theory Using
  Resolution}}.
\newblock {\sl \bibinfo{journal}{J. Autom. Reasoning}}
  \bibinfo{volume}{29}(\bibinfo{number}{3-4}), pp. \bibinfo{pages}{253--275},
  \doi{10.1007/1072195i\_10}.

\bibitemdeclare{incollection}{equality-handbook}
\bibitem{equality-handbook}
\bibinfo{author}{Anatoli Degtyarev} \& \bibinfo{author}{Andrei Voronkov}
  (\bibinfo{year}{2001}): \emph{\bibinfo{title}{Equality Reasoning in
  Sequent-Based Calculi}}.
\newblock In \bibinfo{editor}{John~Alan Robinson} \& \bibinfo{editor}{Andrei
  Voronkov}, editors: {\sl \bibinfo{booktitle}{Handbook of Automated
  Reasoning}}. \bibinfo{publisher}{Elsevier and MIT Press}, pp.
  \bibinfo{pages}{611--706}, \doi{10.1016/B978-044450813-3/50012-6}.

\bibitemdeclare{article}{Dershowitz82}
\bibitem{Dershowitz82}
\bibinfo{author}{Nachum Dershowitz} (\bibinfo{year}{1982}):
  \emph{\bibinfo{title}{Orderings for Term-Rewriting Systems}}.
\newblock {\sl \bibinfo{journal}{Theor. Comput. Sci.}} \bibinfo{volume}{17},
  pp. \bibinfo{pages}{279--301}, \doi{10.1016/0304-3975(82)90026-3}.

\bibitemdeclare{inproceedings}{assoccommut}
\bibitem{assoccommut}
\bibinfo{author}{Nachum Dershowitz}, \bibinfo{author}{Jieh Hsiang},
  \bibinfo{author}{N.~Alan Josephson} \& \bibinfo{author}{David~A. Plaisted}
  (\bibinfo{year}{1983}): \emph{\bibinfo{title}{Associative-Commutative
  Rewriting}}.
\newblock In: {\sl \bibinfo{booktitle}{IJCAI}}. pp. \bibinfo{pages}{940--944}.

\bibitemdeclare{article}{modulo}
\bibitem{modulo}
\bibinfo{author}{Gilles Dowek}, \bibinfo{author}{Th{\'e}r{\`e}se Hardin} \&
  \bibinfo{author}{Claude Kirchner} (\bibinfo{year}{2003}):
  \emph{\bibinfo{title}{Theorem Proving Modulo}}.
\newblock {\sl \bibinfo{journal}{J. Autom. Reasoning}}
  \bibinfo{volume}{31}(\bibinfo{number}{1}), pp. \bibinfo{pages}{33--72},
  \doi{10.1007/3-540-46508-1\_1}.

\bibitemdeclare{article}{codedcontexttrees}
\bibitem{codedcontexttrees}
\bibinfo{author}{Harald Ganzinger}, \bibinfo{author}{Robert Nieuwenhuis} \&
  \bibinfo{author}{Pilar Nivela} (\bibinfo{year}{2004}):
  \emph{\bibinfo{title}{Fast Term Indexing with Coded Context Trees}}.
\newblock {\sl \bibinfo{journal}{J. Autom. Reasoning}}
  \bibinfo{volume}{32}(\bibinfo{number}{2}), pp. \bibinfo{pages}{103--120},
  \doi{10.1023/B:JARS.0000029963.64213.ac}.

\bibitemdeclare{inproceedings}{Graf95}
\bibitem{Graf95}
\bibinfo{author}{Peter Graf} (\bibinfo{year}{1995}):
  \emph{\bibinfo{title}{Substitution Tree Indexing}}.
\newblock In \bibinfo{editor}{Jieh Hsiang}, editor: {\sl
  \bibinfo{booktitle}{Rewriting Techniques and Applications, 6th International
  Conference, RTA-95, Kaiserslautern, Germany, April 5-7, 1995, Proceedings}}.
  {\sl \bibinfo{series}{Lecture Notes in Computer Science}}
  \bibinfo{volume}{914}, \bibinfo{publisher}{Springer}, pp.
  \bibinfo{pages}{117--131}.

\bibitemdeclare{book}{harrison-book}
\bibitem{harrison-book}
\bibinfo{author}{John Harrison} (\bibinfo{year}{2009}):
  \emph{\bibinfo{title}{Handbook of Practical Logic and Automated Reasoning}}.
\newblock \bibinfo{publisher}{Cambridge University Press}.

\bibitemdeclare{inproceedings}{Hurd99}
\bibitem{Hurd99}
\bibinfo{author}{Joe Hurd} (\bibinfo{year}{1999}):
  \emph{\bibinfo{title}{Integrating Gandalf and HOL}}.
\newblock In \bibinfo{editor}{Yves Bertot}, \bibinfo{editor}{Gilles Dowek},
  \bibinfo{editor}{Andr{\'e} Hirschowitz}, \bibinfo{editor}{C.~Paulin} \&
  \bibinfo{editor}{Laurent Th{\'e}ry}, editors: {\sl
  \bibinfo{booktitle}{TPHOLs}}. {\sl \bibinfo{series}{Lecture Notes in Computer
  Science}} \bibinfo{volume}{1690}, \bibinfo{publisher}{Springer}, pp.
  \bibinfo{pages}{311--322}, \doi{10.1007/3-540-48256-3\_21}.
\newblock
  \urlprefix\url{http://link.springer.de/link/service/series/0558/bibs/1690/16%
900311.htm}.

\bibitemdeclare{techreport}{metis}
\bibitem{metis}
\bibinfo{author}{Joe Hurd} (\bibinfo{year}{2003}):
  \emph{\bibinfo{title}{First-Order Proof Tactics in Higher-Order Logic Theorem
  Provers}}.
\newblock \bibinfo{type}{Technical Report}
  \bibinfo{number}{NASA/CP-2003-212448}, \bibinfo{institution}{Nasa technical
  reports}.

\bibitemdeclare{article}{Knuth-Bendix}
\bibitem{Knuth-Bendix}
\bibinfo{author}{Donald Knuth} \& \bibinfo{author}{P.~Bendix}
  (\bibinfo{year}{1970}): \emph{\bibinfo{title}{Simple word problems in
  universal algebras}}.
\newblock {\sl \bibinfo{journal}{Computational problems in Abstract Algebra (J.
  Leech ed.)}} , pp. \bibinfo{pages}{263--297}.

\bibitemdeclare{article}{McCune}
\bibitem{McCune}
\bibinfo{author}{W.~McCune} (\bibinfo{year}{1992}):
  \emph{\bibinfo{title}{Experiments with discrimination tree indexing and path
  indexing for term retrieval}}.
\newblock {\sl \bibinfo{journal}{Journal of Automated Reasoning}}
  \bibinfo{volume}{9(2)}, pp. \bibinfo{pages}{147--167},
  \doi{10.1007/BF00245458}.

\bibitemdeclare{article}{MP08}
\bibitem{MP08}
\bibinfo{author}{Jia Meng} \& \bibinfo{author}{Lawrence~C. Paulson}
  (\bibinfo{year}{2008}): \emph{\bibinfo{title}{Translating Higher-Order
  Clauses to First-Order Clauses}}.
\newblock {\sl \bibinfo{journal}{J. Autom. Reasoning}}
  \bibinfo{volume}{40}(\bibinfo{number}{1}), pp. \bibinfo{pages}{35--60},
  \doi{10.1007/s10817-007-9085-y}.

\bibitemdeclare{article}{MQP06}
\bibitem{MQP06}
\bibinfo{author}{Jia Meng}, \bibinfo{author}{Claire Quigley} \&
  \bibinfo{author}{Lawrence~C. Paulson} (\bibinfo{year}{2006}):
  \emph{\bibinfo{title}{Automation for interactive proof: First prototype}}.
\newblock {\sl \bibinfo{journal}{Inf. Comput.}}
  \bibinfo{volume}{204}(\bibinfo{number}{10}), pp. \bibinfo{pages}{1575--1596},
  \doi{10.1016/j.ic.2005.05.010}.

\bibitemdeclare{incollection}{paramodulation}
\bibitem{paramodulation}
\bibinfo{author}{Robert Nieuwenhuis} \& \bibinfo{author}{Alberto Rubio}
  (\bibinfo{year}{2001}): \emph{\bibinfo{title}{Paramodulation-based thorem
  proving}}.
\newblock In \bibinfo{editor}{John~Alan Robinson} \& \bibinfo{editor}{Andrei
  Voronkov}, editors: {\sl \bibinfo{booktitle}{Handbook of Automated
  Reasoning}}. \bibinfo{publisher}{Elsevier and MIT Press}, pp.
  \bibinfo{pages}{471--443}, \doi{10.1016/B978-044450813-3/50009-6}.
\newblock \bibinfo{note}{ISBN-0-262-18223-8}.

\bibitemdeclare{article}{blast}
\bibitem{blast}
\bibinfo{author}{Lawrence~C. Paulson} (\bibinfo{year}{1999}):
  \emph{\bibinfo{title}{A Generic Tableau Prover and its Integration with
  Isabelle}}.
\newblock {\sl \bibinfo{journal}{J. UCS}}
  \bibinfo{volume}{5}(\bibinfo{number}{3}), pp. \bibinfo{pages}{73--87}.

\bibitemdeclare{article}{vampire_annals}
\bibitem{vampire_annals}
\bibinfo{author}{Alexandre Riazanov} \& \bibinfo{author}{Andrei Voronkov}
  (\bibinfo{year}{2002}): \emph{\bibinfo{title}{The Design and Implementation
  of Vampire}}.
\newblock {\sl \bibinfo{journal}{AI Communications}} \bibinfo{volume}{15(2-3)},
  pp. \bibinfo{pages}{91--110}.

\bibitemdeclare{article}{LRS}
\bibitem{LRS}
\bibinfo{author}{Alexandre Riazanov} \& \bibinfo{author}{Andrei Voronkov}
  (\bibinfo{year}{2003}): \emph{\bibinfo{title}{Limited resource strategy in
  resolution theorem proving}}.
\newblock {\sl \bibinfo{journal}{J. Symb. Comput.}}
  \bibinfo{volume}{36}(\bibinfo{number}{1-2}), pp. \bibinfo{pages}{101--115},
  \doi{10.1016/S0747-7171(03)00040-3}.

\bibitemdeclare{article}{setoidrewriting}
\bibitem{setoidrewriting}
\bibinfo{author}{Matthieu Sozeau} (\bibinfo{year}{2009}):
  \emph{\bibinfo{title}{A New Look at Generalized Rewriting in Type Theory}}.
\newblock {\sl \bibinfo{journal}{Journal of Formalized Reasoning}}
  \bibinfo{volume}{2}(\bibinfo{number}{1}), pp. \bibinfo{pages}{41--62}.

\bibitemdeclare{article}{Sutcliffe09}
\bibitem{Sutcliffe09}
\bibinfo{author}{Geoff Sutcliffe} (\bibinfo{year}{2009}):
  \emph{\bibinfo{title}{The 4th IJCAR Automated Theorem Proving System
  Competition - CASC-J4}}.
\newblock {\sl \bibinfo{journal}{AI Commun.}}
  \bibinfo{volume}{22}(\bibinfo{number}{1}), pp. \bibinfo{pages}{59--72}.

\bibitemdeclare{article}{demodulation}
\bibitem{demodulation}
\bibinfo{author}{Larry Wos}, \bibinfo{author}{George~A. Robinson},
  \bibinfo{author}{Daniel~F. Carson} \& \bibinfo{author}{Leon Shalla}
  (\bibinfo{year}{1967}): \emph{\bibinfo{title}{The Concept of Demodulation in
  Theorem Proving}}.
\newblock {\sl \bibinfo{journal}{J. ACM}}
  \bibinfo{volume}{14}(\bibinfo{number}{4}), pp. \bibinfo{pages}{698--709},
  \doi{10.1145/321420.321429}.

\end{thebibliography}

\end{document}